\documentclass{article}
\usepackage{spconf,amsmath,graphicx}
\usepackage{amsfonts}
\usepackage{epsfig}
\usepackage{epstopdf}
\usepackage{color}       

\usepackage{tikz}
\usetikzlibrary{snakes,arrows,shapes}

\newcommand{\R}{\mathbb R}

\newcommand{\N}{{\mathbb N}}

\newcommand{\comment}[1]{}
\newcommand{\Xv}{\underline{X}}

\graphicspath{
{Figures/}
}

\title{Matrix Products for the Synthesis of Stationary Time Series \\ with a priori Prescribed Joint Distributions}
\name{Florian Angeletti, Eric Bertin,  Patrice Abry}
\address{
Physics Dept., CNRS, UMR5672,  ENS de Lyon, France.\\
{\tt ens-lyon.fr/PHYSIQUE/}, {\tt firstname.lastname@ens-lyon.fr}
}

\def\R{\mathbb{R}}
\def\N{\mathbb{N}}
\def\Prob{\mathbb{P}}

\def\Esp#1{\mathbb{E}\left[#1\right]}

\newcommand{\Tr}[1]{\mathsf{tr} \left( #1 \right) }

\newcommand{\E}{\mathcal{E}}

\newcommand{\Rd}{R_d}
\newcommand{\Lf}{\mathcal{L}}

\newcommand{\Mq}[1]{M\!\left(#1\right)}

\newcommand{\MP}{\mathcal{P}}

\newcommand {\GamF} [3] {#1_{\Gamma_{#2}, \Gamma_{#3} } }

\newcommand{\F}{\mathcal{F} }
\newcommand{\Normal}[2]{\mathcal{N} _{ #1,#2 } }
\newcommand{\Real}[1]{\mathfrak{Re}\left\{ #1 \right\}}

\begin{document}
%
\maketitle
\begin{abstract}
Inspired from non-equilibrium statistical physics models, a general framework enabling the definition and synthesis of stationary time series with a priori prescribed and controlled joint distributions is constructed. 
Its central feature consists of preserving for the joint distribution the \emph{simple} product structure it has under independence while enabling to input controlled and prescribed dependencies amongst samples. 
To that end, it is based on products of $d$-dimensional matrices, whose entries consist of valid distributions. 
The statistical properties of the thus defined time series are studied in details. 
Having been able to recast this framework into that of Hidden Markov Models enabled us to obtain an efficient synthesis procedure. 
Pedagogical well-chosen examples (time series with the same marginal distribution, same covariance function, but different joint distributions) aim at illustrating the power and potential of the approach and at showing how targeted statistical properties can be actually prescribed.
\end{abstract}
\section{Introduction}
\label{sec:intro}

In modern signal processing, the need to produce numerically random vectors whose joint distributions are fully prescribed and controlled is more and more common.
Typical examples are Bayesian estimation (with Monte Carlo Markov Chain or variational based resolution schemes, cf.~e.g., \cite{tlg2008}) or estimation performance assessment (where estimators are benchmarked on time series with prescribed distributions). 
For univariate Gaussian stationary time series, the so-called \emph{Circulant Embedded Matrix} synthesis procedure \cite{DietrichNewsam1997} is considered as the state of the art solution. 
For non Gaussian stationary time series, various approaches were proposed that aim at controlling both the marginal distribution and the covariance structure of the process (cf. e.g., \cite{GrigoriuBook,hpa10} and references therein for reviews). 
Often, such methods suffer from two major limitations: The joint distribution cannot be prescribed a priori and is hence not controlled and results as a consequence of the details of the synthesis procedure; Distributions that consist of mixtures of elementary laws cannot always easily be combined with prescribed covariance. 
The very general framework of Markov Chain simulation offers an alternative and broad class of solutions,
focusing on the modeling of local dynamical properties, while not explicitly putting the emphasis on a direct prescription of the joint distributions of the process. 

Instead, in the present contribution, the focus is on the synthesis of stationary time series whose joint distributions are explicitly prescribed and chosen a priori. 
It is based on a construction inspired from out-of-equilibrium statistical physics models (cf. e.g., \cite{DerridaASEP93,Mallick97,Evans2007}). It is founded on the central idea that the joint distribution of a vector $\underline{X}_N={x_1, x_2, \ldots, x_N} $ is still written as a (oriented) product:  $ p_{\underline{X}} \propto \prod_{k=1}^{N} R_d(x_k)$, where, however the $R_d$s are not single distributions, but instead $d$-dimensional matrices of distributions.
The key results of the present contribution consist of first deriving the general properties of this construction and then to specify it to a very interesting case specifically suited for stationary time series with all joint distributions prescribed (cf. Section 2). 
An efficient synthesis procedure is devised by explicitly recasting this framework into that of Hidden Markov Models (cf. Section 3). 
Finally, explicit examples are studied and simulated numerically, aiming at illustrating the power of the proposed approach: They consist of time series sharing the same complicated (mixture) marginal distribution, the same covariance functions, but different joint distributions (cf. Section 4). 

\section{Joint distributions as matrix product}


\noindent {\bf General Framework.} \quad  Let us first define a general formalism for the construction of joint distribution functions of random vectors $\underline{X}_N$, inspired from non equilibrium statistical physics models (cf. e.g. \cite{DerridaASEP93,Mallick97,Evans2007}) and based on products of matrices of distributions.
Let $ \Rd(x)$ denote a $d$-dimensional matrix with entries 
\begin{equation}
\label{equ-Rd} 
\Rd(x)_{i,j}=\E_{i,j} \MP(x)_{i,j},
\end{equation}
where $\MP(x)_{i,j} $ are valid distribution functions and $ \E_{i,j}  $ arbitrary positive numbers, forming the matrix $ \E$. 
Let $A$ denote an arbitrary, but fixed strictly positive and non-random, matrix and $\Lf(M)$ the linear form applied to matrix $M$ defined as $\Lf(M)= \Tr {A^T M}$. 

Further, let $\Xv \equiv \{X_n\}_{1\le n \le N}$ denote a random vector, of chosen size $N$, explicitly defined via its joint distribution: 
\begin{equation} 
\label{eqn:prob}
\Prob(x_1,\dots,x_N)=\frac{\Lf \left( \prod_k^N \Rd(x_k) \right) }{\Lf \left( \E^N \right) },  
\end{equation}
where $\prod_k^N \Rd(x_k)=R_d(x_1) \ldots R_d(x_N)$ denotes the oriented product (i.e., the order of the factors is fixed and cannot be changed).
It is straightforward to check that the joint choice of strictly positive entries for matrices $A$ and $\E$, and of valid distribution functions for $\MP(x)_{i,j} $ is sufficient to ensure that Eq. (\ref{eqn:prob}) defines a valid joint distribution function.

From these definitions, calculations and matrix manipulations not reported here enabled us to derive a number of statistical properties of the vector  $\{X_n\}_{1\le n \le N}$. 
Its univariate (marginal) distributions and one-sample moments take explicit forms (with $\Mq{q}  = \int_\R x^q \Rd(x) dx$): 
\begin{eqnarray} 
\Prob(X_k=x) &= &\frac{\Lf \left( \E^{k-1}\Rd(x)\E^{N-k} \right) }{\Lf \left(\E^N \right)}, \\
\Esp{X_k^q} &= & \frac{\Lf ( \E^{k-1}\Mq{q}\E^{N-k})}{\Lf(\E^N)}.
\end{eqnarray}
Furthermore, the joint $p$-sample moments read (with $p\in N$, $k_1<\dots<k_p$ and $q_r $ the order associated to the entry $x_{k_r}$): $ \Esp{\prod_{r=1}^p X_{k_r}^{q_r}}= $
\begin{equation} 
 \frac{\Lf \left(\E^{k_1-1} \left( \prod_r^{p-1} \Mq{q_r} \E^{k_{r+1}-k_{r}-1}  \right) \Mq{q_s} \E^{N-k_{s}} \right) }{\Lf(\E^N)} 
\end{equation}


\noindent {\bf Stationary time series.} \quad Let us now focus on specific choices for  matrices $A$ and $E$, of interest here to construct stationary time series: 
\begin{equation}
A_{i,j}=  \frac{1}{d}, \quad \E =  \alpha I_d + \beta J_d, \makebox{ with } \alpha+\beta=1, 
\end{equation} 
$I_d$ is the $d$-dimensional unity matrix and $J_d \in M_d(\R) $ defined as: 
\comment{
\[ J_d = 
\begin{pmatrix}
0       & 1      & 0     \\
        & \ddots & \ddots 1\\
1       &        & 0  \\
\end{pmatrix}
\]
}

\[ J_d = 
\begin{pmatrix}
0       & 1      &0       & \cdots  & 0      \\
\vdots  & \ddots & \ddots & \ddots  & \vdots \\
\vdots  &        & \ddots & \ddots  & 0      \\
0       &        &        & \ddots  & 1      \\
1       & 0      & \cdots & \cdots  & 0 \\  
\end{pmatrix}
\]

Such choices for $A$ and $\E$ enabled us to show (full calculations not reported here) that: 
\begin{eqnarray} 
\forall \E , n \in \N & \Lf(\E^n) & = 1,\\
\forall M, \E, (n,r) \in \N^2, & \frac{\Lf(M E^r)} {\Lf(E^n)} & = \Lf(M).
\end{eqnarray}
This further enables us to obtain  that the results above can be specified as: 
\begin{equation} 
\Prob(X_k=x) = \frac{1}{d} \sum_{i,j} \E_{i,j} \MP_{i,j} (x), 
\end{equation}
\begin{equation} 
\Esp{X_k^q} =  \Lf \left( \Mq{q} \right), 
\end{equation}
\begin{equation} 
\Esp{\prod_{r=1}^p X_{k_r}^{q_r}} =  \Lf \left( \left( \prod_r^{p-1} \Mq{q_r} \E^{k_{r+1}-k_{r}-1}  \right) \Mq{q_p}  \right). 
\label{equ-Mq}
\end{equation}

These relations clearly indicate that the vector $\{ X_n \}_{1\le n \le N}$ can now be regarded as a stationary time series: All joint statistics depend only on time differences, $k_{r+1}-k_{r}$.
Both its marginal and joint probabilities are prescribed by the choices of $\E$ and of the $  \MP_{i,j}$.
Also, Eq. (\ref{equ-Mq}) constitutes a key result with respect to applications as it clearly shows that the joint statistics of order $q$ of the time series can be prescribed by the sole selection of a suitable $M(q)$ matrix. 
This will be explicitly used in Section 4.

The form of the covariance function can be further studied. 
The eigenvalues of $\E$ read $ \lambda_k= \alpha + \beta e^{\frac{2 \imath \pi k}{d}} $ and can be rewritten as $ \lambda_k= e^{-\frac{1}{\tau_k}} e^{\pm \imath \frac{2 \pi}{T_k} } $,  $k=1, \ldots \lfloor d/2 \rfloor$ (where $\lfloor z \rfloor $ stands for the integer part of $z$). 
For ease of notations, let  $ (C_M)_k = \sum_l M_{l,k}$,  $(L_M)_k= \sum_c M_{k,c} $ and let $\F$ denote the (non-normalized) discrete Fourier transform. 
Then, detailed calculations enabled us to show that, for any $q$, 
\begin{equation}
\begin{aligned}
\label{eqn:eigen:fft}
&\Esp {X_0 ^q X_t^q} -\Esp{X_0^q} \Esp{X_t^q}=\\
&\sum_{k=1}^{\lfloor d/2 \rfloor }  m_k \Real  { \F(L_{\Mq{q}})_k  \overline{\F(C_{\Mq{q}})_k } e^{-\frac{t-1}{\tau_k}} e^{ \imath \frac{2 \pi (t-1)}{T_k} }}, 
\end{aligned} 
 \end{equation}
 where $m_k = 1 $ if $ 2k=d$ and $2$ otherwise. 
 This result shows that $\tau_k$ and $T_k$ are characteristic dependence lengths and periodicities that depend on the joint choices of $d$ and $ \alpha $. More precisely, from the definition of  $ \tau_k=- 1/ \ln  |\lambda_k|$, we can show that $\tau_k \approx_{\alpha \beta \rightarrow 0} \left[ \alpha \beta \left(1- \cos \frac{2 \pi k}{d} \right) \right]^{-1}$.
Enlarging the dimension $d$ of the matrices $\Rd$ increases both the number of distinct dependence lengths as well as the ratio of the smallest to the largest such characteristic lengths, which varies asymptotically as $(d/2\pi)^2/2$. 

Therefore, choosing $d$, $\alpha$, hence $\E$, and the sequence of $\MP_{i,j} $ enables to select the marginal and joint distributions according to given targets. 
Section 4 will illustrate the potential of the method. 

\section{Synthesis: Hidden Markov Chain}

While the compact form of $\Prob(\Xv)$ in Eq.~(\ref{eqn:prob}) constitutes a general framework for the analytical derivation of numerous statistical properties of $\Xv$, it provides few insights with respect to its numerical synthesis. 
To address this issue, a reformulation of Eq.~(\ref{eqn:prob}) into the form of a Hidden Markov Model is now devised. 

First, extending to product of $N$ matrices the fact that the entries of the matrix product $(ABC)$ reads $(ABC)_{i,j} = \sum_{k,l} a_{i,k}b_{k,l} c_{l,j}$, we have been able to recast  Eq.~(\ref{eqn:prob}) into:
\begin{equation}
\label{eqn:synth_form}
 \Prob(x_1,\ldots,x_N)=  \sum_{\underline \Gamma }   \kappa(\underline{\Gamma}) \prod_{k=1}^{N} \GamF{\MP(x_k)}{k-1}{k} , 
\end{equation} 
with $ \underline{\Gamma} \equiv \{\Gamma_0, \ldots,\Gamma_k, \ldots, \Gamma_N \} \in [1,\ldots,d]^{N+1}$ and
\begin{equation} 
\label{eqn:kappa}
\kappa (\underline{\Gamma})= \frac{ \GamF{A}{0}{N}}{\Lf (\E^N)} \prod_{k=1}^N \GamF{\E}{k-1}{k}. 
\end{equation} 
It is straightforward to verify that $\sum_{\underline{\Gamma}} \kappa (\underline{\Gamma})=1$, hence $\kappa (\underline{\Gamma})$ can be interpreted as the probability function of $\underline{\Gamma}$. 
Moreover, Eq.~(\ref{eqn:synth_form}) shows that $\Prob(\Xv)$ can be read as a $ \kappa (\underline{\Gamma})$-weighted mixture 
of laws, each defined as the product $\prod_{k=1}^N \GamF{\MP(x_k)}{k-1}{k+1} $. 

\begin{figure}
\begin{center}
\includegraphics[height=2cm]{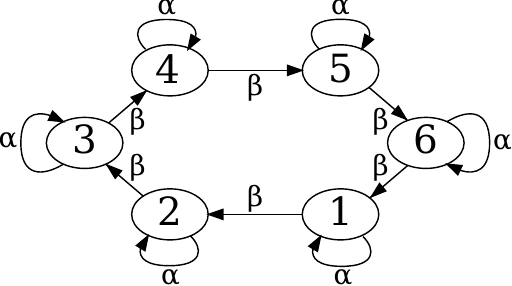}
\end{center}
 \vspace{-0.5cm}
\caption{\label{Fig:markov} {\bf Transition graph.} For $d= 6 $ and $\E= \alpha I_d + \beta J_d$ }
\end{figure}

Second, from Eq. (\ref{eqn:kappa}), using the special form of $A$ and $\E$, it can be shown that: 
\begin{equation}
\begin{split}
 \Prob(\Gamma_{k}=j | \Gamma_{0}=\gamma_0,\ldots \Gamma_{k-1}=\gamma_{k-1}) = \\
  \frac{\E_{\gamma_0,\gamma_{1}} \ldots \E_{\gamma_{k-1}, j}  \sum_\Gamma  \E_{j,\Gamma_{k+1}}  \ldots \E_{\Gamma_{N-1},\Gamma_N} } 
  {\E_{\gamma_0,\gamma_{1}} \ldots \E_{\gamma_{k-2},\gamma_{k-1}} \sum_\Gamma  \E_{\gamma_{k-1}, \Gamma_{k}}  \ldots \E_{\Gamma_{N-1},\Gamma_{N} } } \\ 
 			              =   \E_{\gamma_{k-1},j}   
 			               = \Prob(\Gamma_{k}=j | \Gamma_{k-1}=\gamma_{k-1})  
\end{split}
\end{equation}
and hence that $ \underline{\Gamma}$ is a homogeneous $d$-state Markov chain, with transition probability matrix: 
\begin{equation}
\label{eqn:transitions}
\Prob(\Gamma_{k+1}=j | \Gamma_k=i)= \E_{i,j}.
\end{equation}
Hence, $\E$ can be recast as the transition matrix of the Markov chain $\underline{\Gamma}$, as illustrated in Fig.~\ref{Fig:markov}.

Third, the final step required for numerical synthesis is to derive the initial distribution for $\Gamma_0$. 
Eq.~(\ref{eqn:kappa}) enabled us to show that it should follow a uniform distribution:
\begin{equation}
\label{eqn:initial_states}
\Prob(\Gamma_{0}=i)= \frac{1} {d}.
\end{equation}

Therefore, the time series $\Xv_k$, as defined from Eq.~(\ref{equ-Rd}), can be read and interpreted as a Hidden Markov Model, with $2$ hidden states: 
the current state $\Gamma_k$ and the previous state $\Gamma_{k-1}$. 

Combining Eqs.~(\ref{eqn:transitions}) and (\ref{eqn:initial_states}), a synthesis algorithm can be sketched as follows: \\
Step 1 : Initialization: Use Eq.~(\ref{eqn:initial_states}) to generate the state $\Gamma_0$. \\
Step 2 : Iteration on $k$: \\
\indent i) Choose at random state $\Gamma_k$, using Eq.~(\ref{eqn:transitions}), \\
\indent ii) Generate $X_k$ according to $\MP_{\Gamma_{k-1},\Gamma_{k}}$ .

\section{Illustrations}
\newcommand{\sigA}{\sigma_1}
\newcommand{\sigB}{\sigma_2}
\newcommand{\Yv}{\underline{Y}}

\begin{figure}[htb]
\centerline{\includegraphics[width=0.49\linewidth,height=25mm]{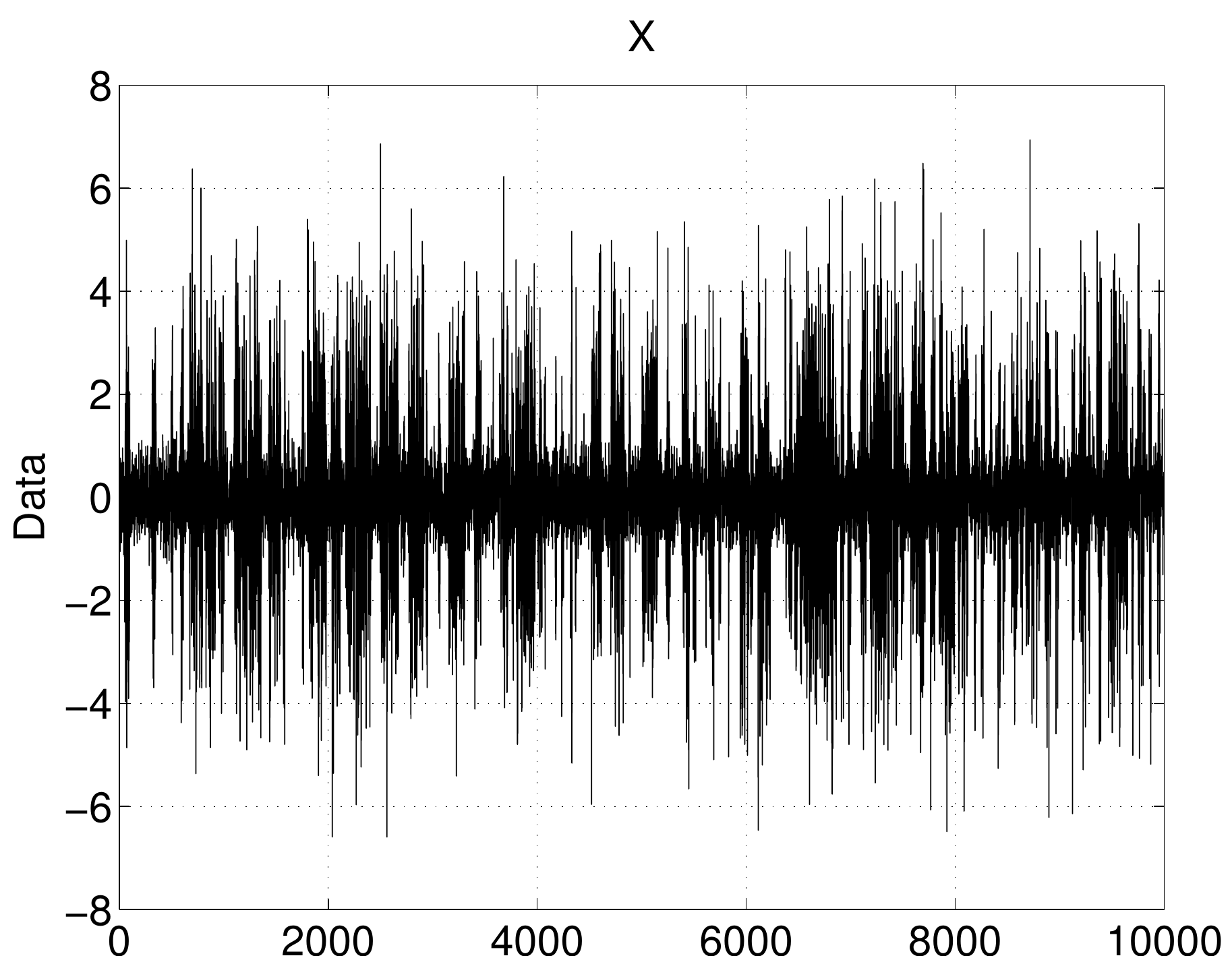} 
\includegraphics[width=0.49\linewidth,height=25mm]{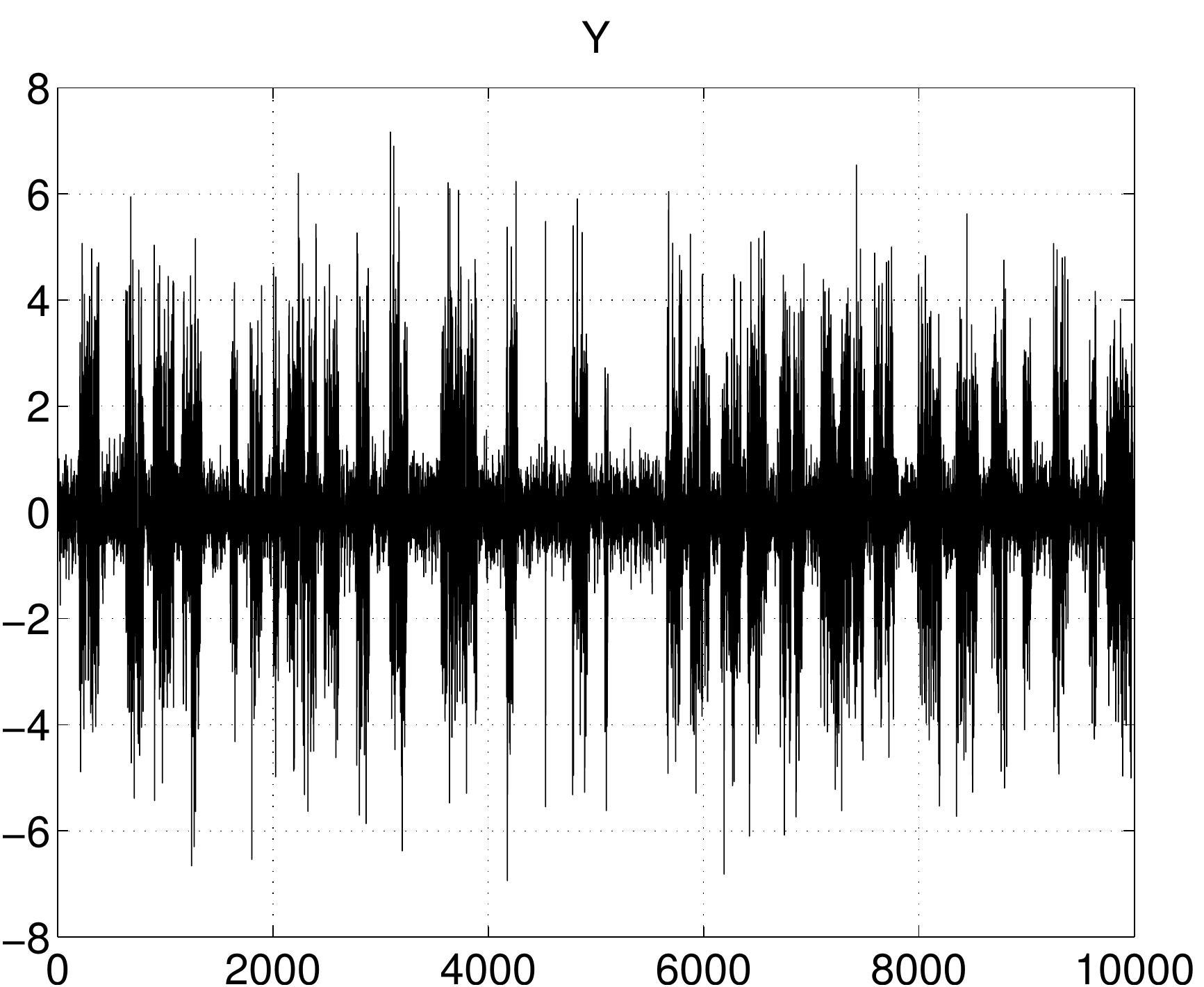}}
\centerline{\includegraphics[width=0.49\linewidth,height=25mm]{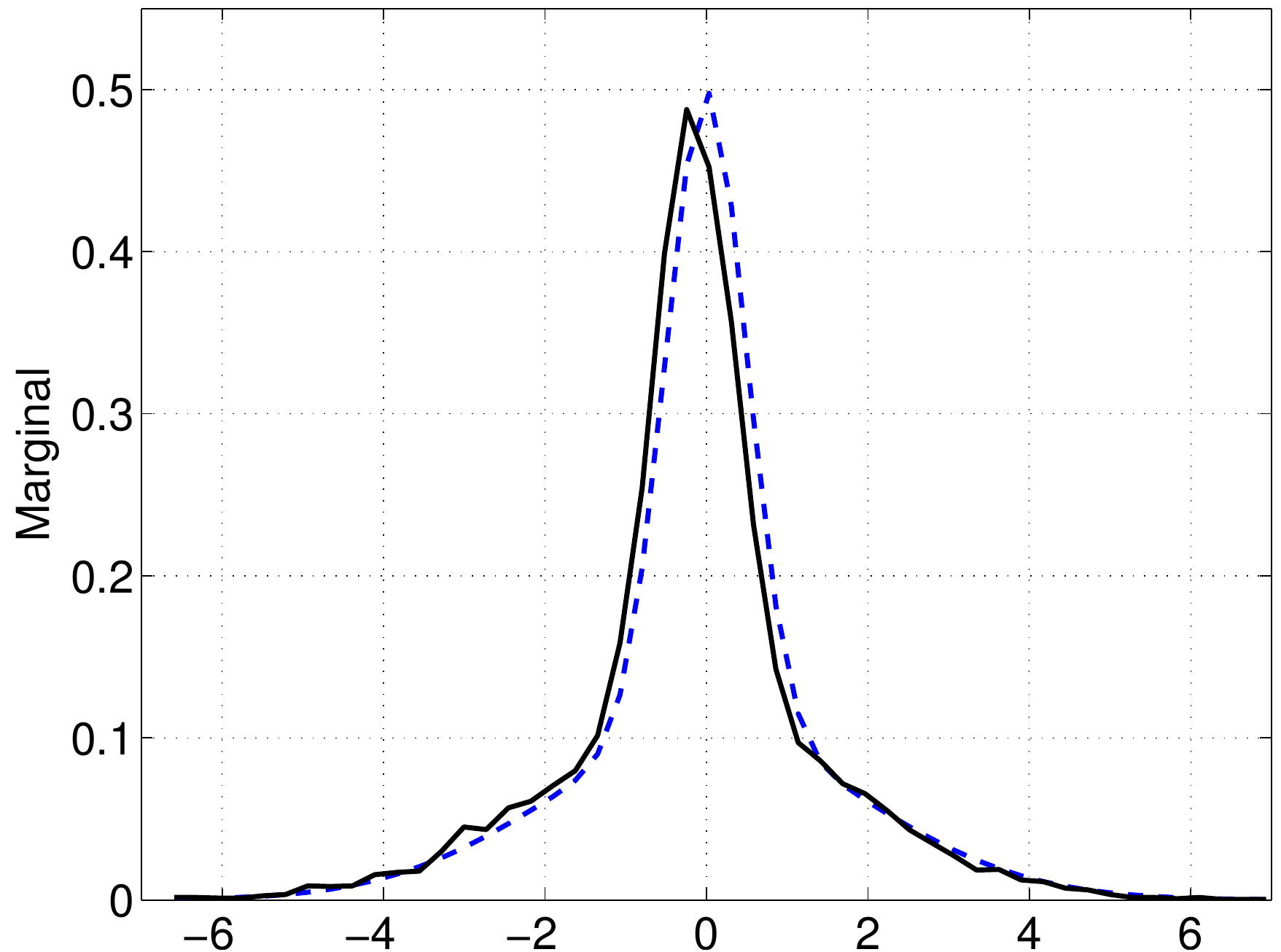}
\includegraphics[width=0.49\linewidth,height=25mm]{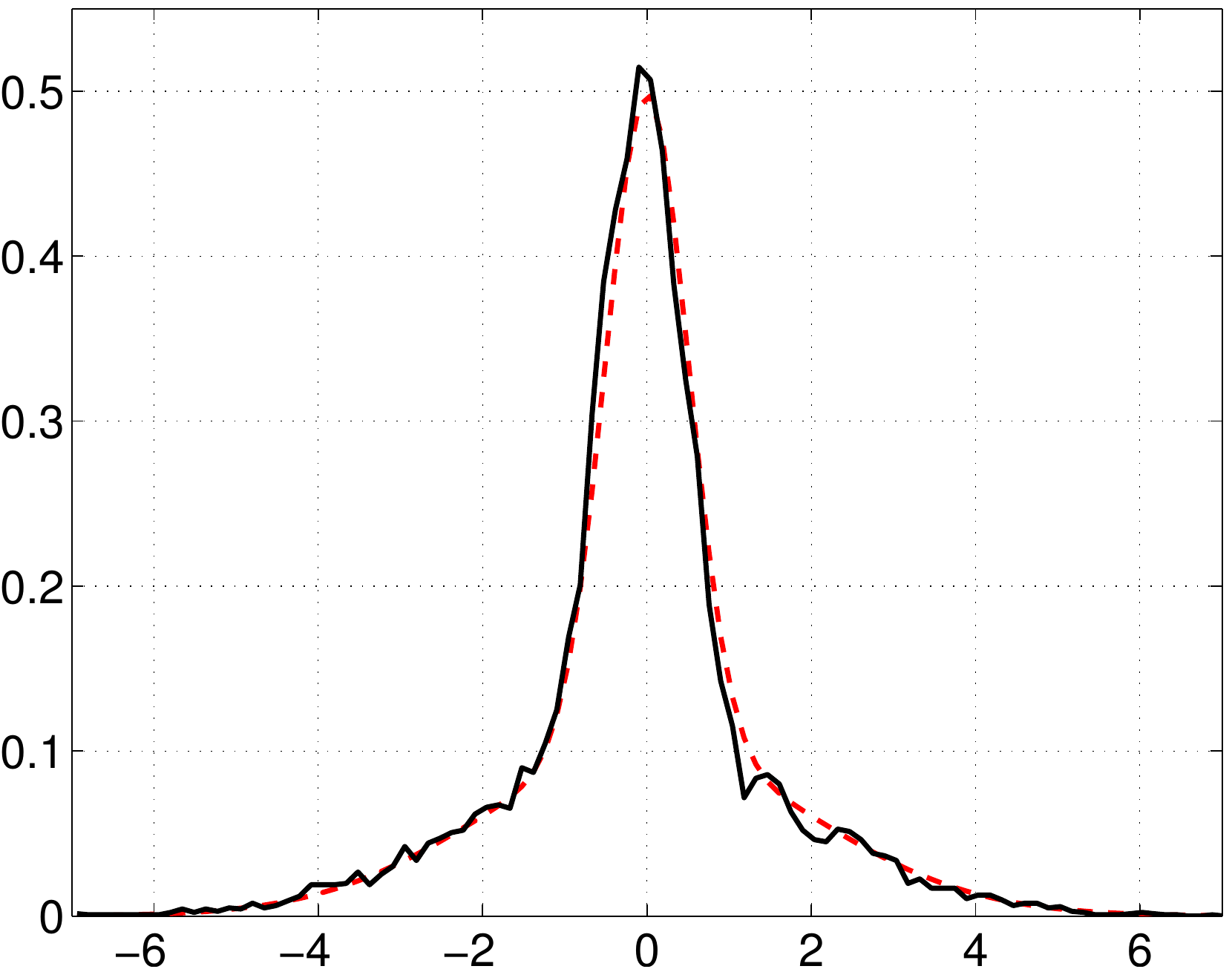}}
\centerline{\includegraphics[width=0.49\linewidth,height=25mm]{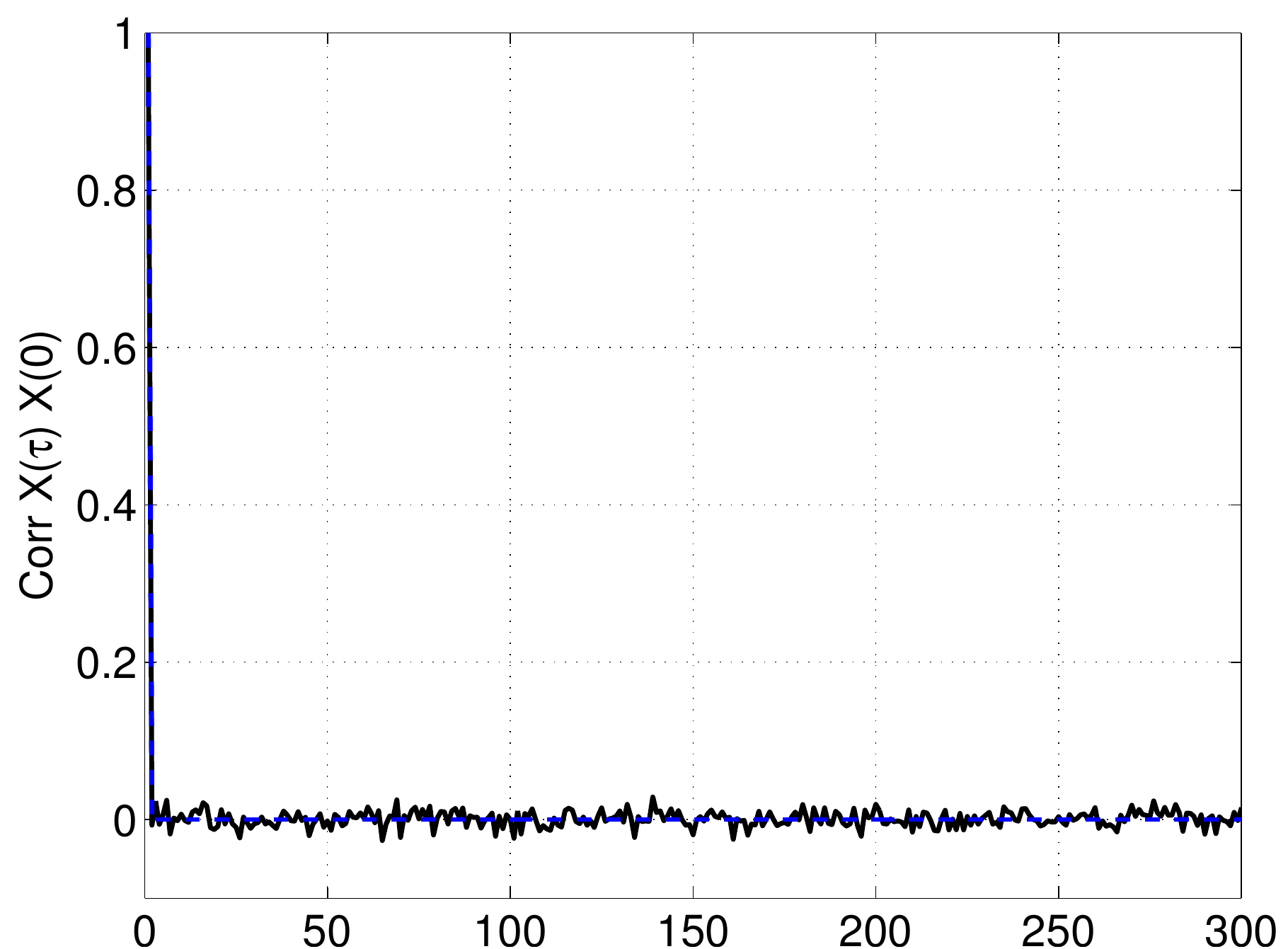}
\includegraphics[width=0.49\linewidth,height=25mm]{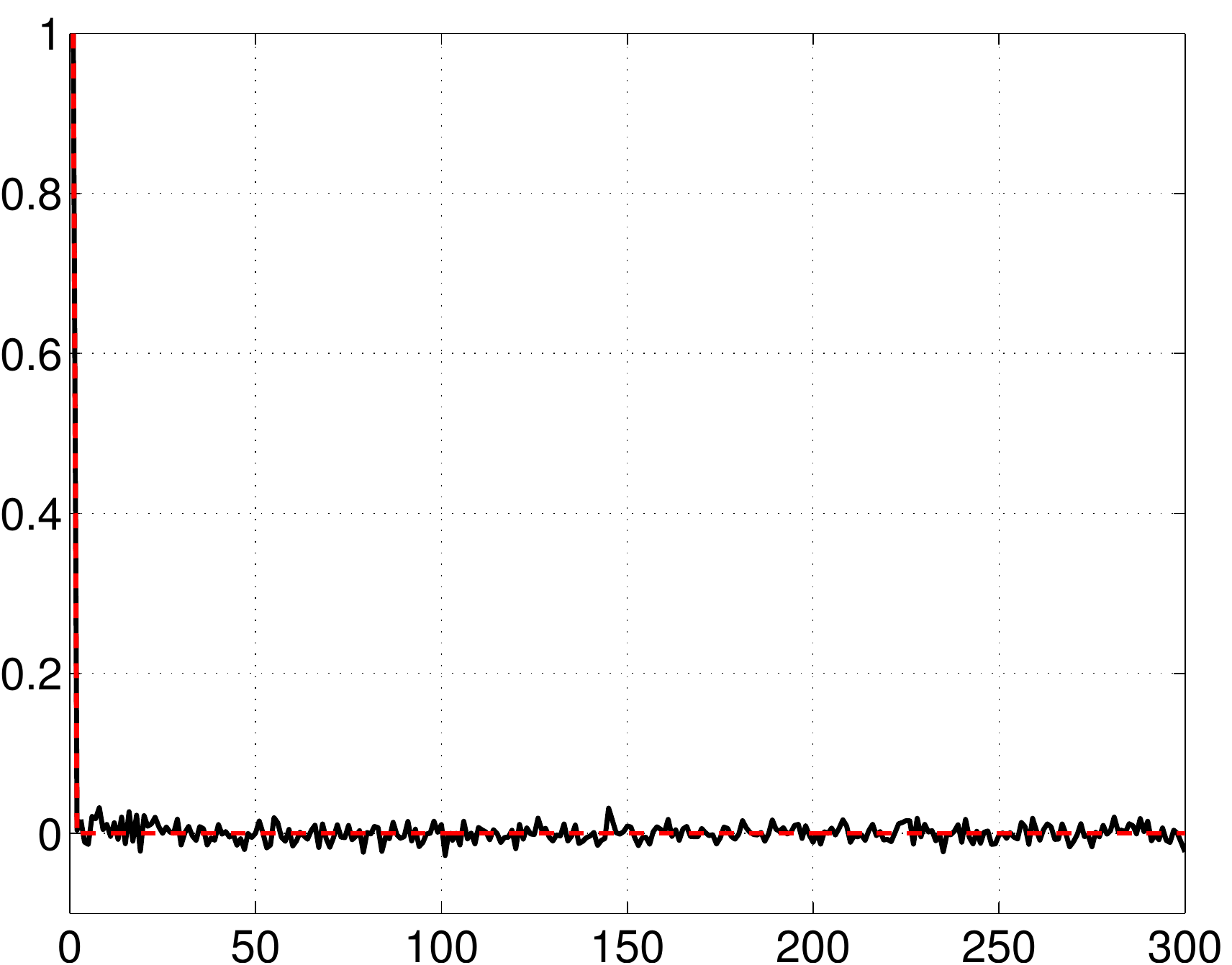}}
\centerline{\includegraphics[width=0.49\linewidth,height=25mm]{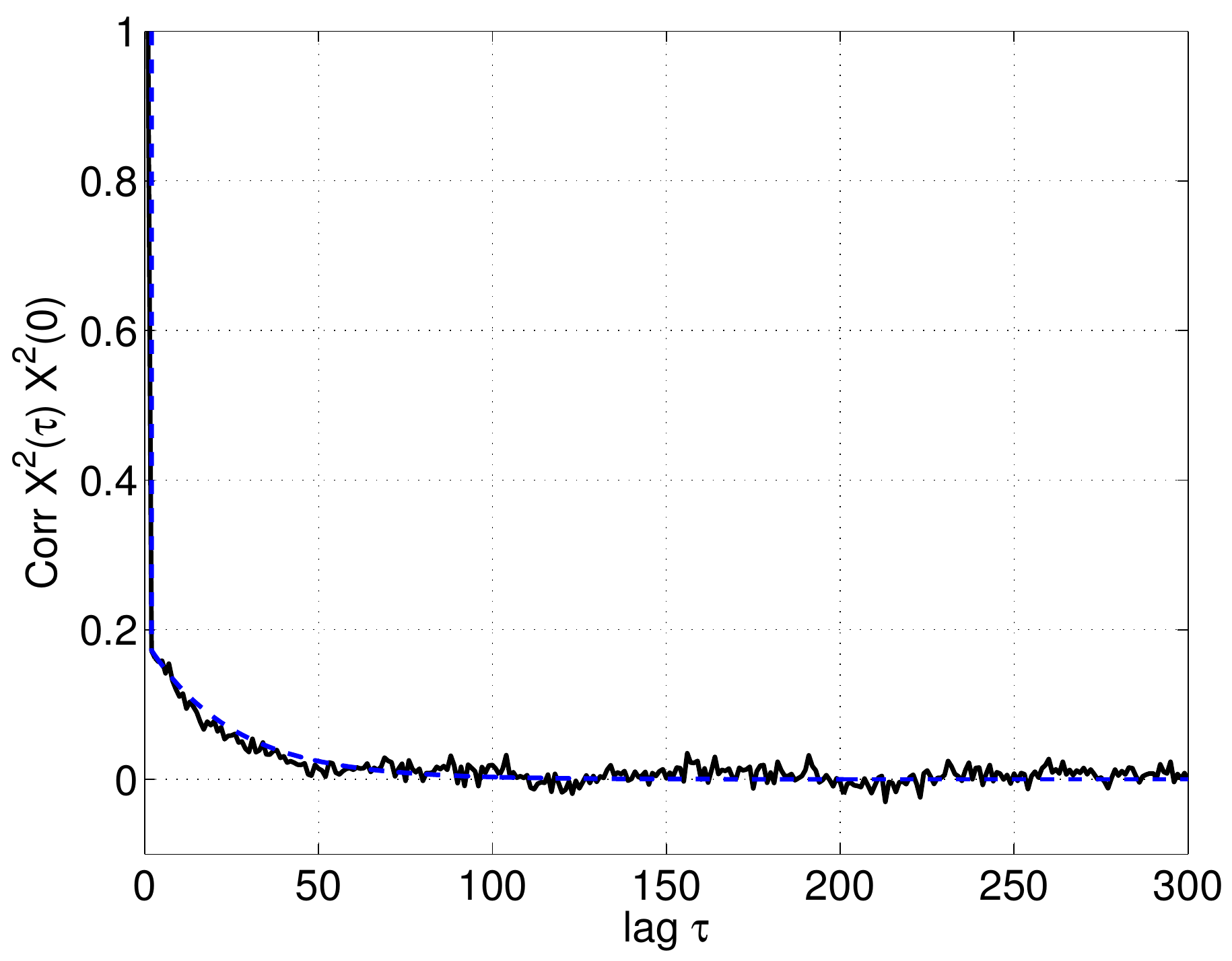}
\includegraphics[width=0.49\linewidth,height=25mm]{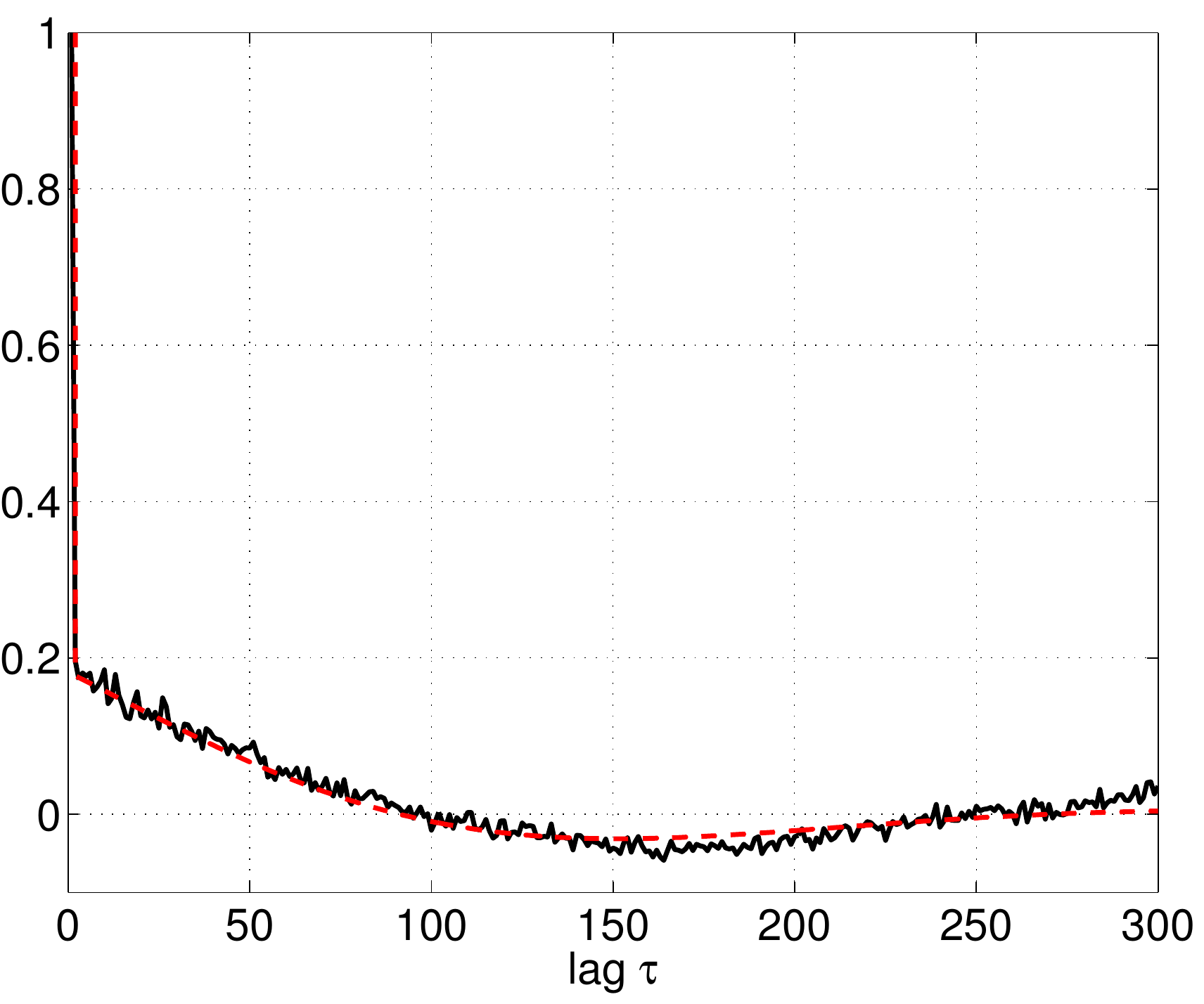}}

\caption{\label{fig:results} {\bf Numerical Synthesis.} Two different times series with the same marginal distribution (mixture of two Gaussians, $\sigA=0.5$, $\sigB=2$), the same covariance functions (chosen as $\delta$ functions) but different covariance functions for their squares, hence different prescribed joint distributions ($\alpha=0.98$). Left side $\Xv$, right side $\Yv$. First line, one realization of the time series. Then, from top to bottom, estimated (solid black lines) and theoretical (dashed colored lines) marginals, correlation functions and correlation functions for the squared time series.}
\end{figure}

To illustrate the potential of the proposed time series theoretical construction and synthesis procedure, a pedagogical example is proposed and the construction of the targeted statistical properties is devised: It consists of a pair $(\Xv,\Yv)$ of processes sharing the same marginal distributions (a mixture of two Gaussian laws), the same autocovariance functions (a $\delta$ function, i.e., no correlation), but different joint distributions, hence different higher order statistic dependences. In this example, the dependence of the $4-$th order statistics will be prescribed. 

We select $d=6$ and $\E= \alpha I_d + \beta J_d$. 
For the sake of simplicity, the univariate (or marginal) distributions is set to be a mixture of two Gaussian distributions: $p(x) = \frac{1}{2} \left( \Normal{0}{\sigA}(x) + \Normal{0}{\sigB}(x)  \right)$.
For that case, it is natural to envisage that the matrix $\MP$ consists only of Gaussian distributions.
Note however that this is not the sole possibility. 
This restriction implies that $\Normal{0}{\sigA}$ and $\Normal{0}{\sigB}$ must appear exactly three times both in the principal and upper circular diagonals of $\MP$. 
In other words, selecting the univariate distribution fixes the number of occurrences of each law in matrix $\MP$, but not their position in the matrix.

Moreover, to  make the example convincing, we chose the covariance functions of both $\Xv$ and $\Yv$ to be $\delta$-functions. 
These two time series will hence have no autocorrelation but (higher-order) dependencies. 
To achieve this, it is sufficient that matrix $\Mq{1}$ is  a zero matrix (cf. Eq.~(\ref{equ-Mq})).
In our example, this is automatically obtained due to the choice of zero-mean distributions: $\Normal{0}{\sigA}$ and $\Normal{0}{\sigB}$. 

These two time series have hence the same marginal distributions and covariance functions. They are yet different by construction as we now impose that they have different joint distributions. 
Taking into account the circularity of the transition graph in Fig.~\ref{Fig:markov}, there exist $80$ distinct choices of $\MP$, consisting of exactly $6$ entries set to  $\Normal{0}{\sigA}$ and $6$ entries set to $\Normal{0}{\sigB}$, leading to different joint distributions. 
Making use of Eq.~(\ref{eqn:eigen:fft}) and of the explicit form of the eigenvalues of $\E$ for $d=6$ enables us to show that $\Esp {X_0 ^q X_t^q} -\Esp{X_0^q} \Esp{X_t^q}$, for any $q$, stems from the superimposition of three distinct exponential terms whose characteristic lengths, we derived asymptotically: $ \tau_1 \underset{\beta \rightarrow 0}{\approx} \frac{2}{ \beta}$, $\tau_2 \underset{\beta\rightarrow 0}{\approx} \frac{3}{2 \beta}$ and $\tau_3 \underset{\beta\rightarrow 0}{\approx} \frac{1}{2 \beta}$. 
The actual choice for the entries of $\MP$ fixes the complex valued coefficients, associated to each of these correlation lengths (cf. Eq.~\ref{eqn:eigen:fft}). 
As demonstrative examples, we selected for $X$: $\MP^X_{i,j}= \Normal{0}{\sigA}$ if $i$ even, and $\MP^X_{i,j}= \Normal{0}{\sigB}$ if $i$ odd. 
Applying Eq.~(\ref{eqn:eigen:fft}) to that choice, shows that the sole $\tau_3$, i.e., the shortest correlation length, appears in the auto-covariance of $\Xv^2$ (the complex-valued coefficients associated to the two other correlation lengths are forced to $0$ by the specific choice made here): $\Esp{X_0^2 X_t^2}-\Esp{X_0^2}\Esp {X_t^2}= \frac{(\sigA -\sigB)^2}{4}(\alpha -\beta)^t$. 
Conversely, choosing $\MP^Y_{i,j}= \Normal{0}{\sigA} \text{if } 1 \le i \le 3 $ and $\MP^Y_{i,j}=\Normal{0}{\sigB} \text{if } 4 \le i \le 6$
leads 
to $\Esp{X_0^2 X_t^2}-\Esp{X_0^2}\Esp {X_t^2}= \frac{(\sigA -\sigB)^2}{d^2} \left[(\alpha-\beta)^{t}+ 4 \Real{(2-\beta-\sqrt{3} \imath \beta) \lambda_1^{t-1} } \right]$.
In that case, both $\tau_1$ and $\tau_3$ contribute to the autocovariance of $\Yv^2$. 
However, $\tau_1 \approx 4 \tau_3$ is dominant at large $t$ and hence constitutes the leading term. 
Using the synthesis procedure devised in Section 3 for this pair of examples yields Fig.~$\ref{fig:results}$, showing for times series $\Xv$ (left column) and $\Yv$ (right column), a particular realization, the estimated and targeted univariate distributions, covariance functions and covariance functions for the squared time series (from top to bottom). 
It clearly shows that $\Xv$ and $\Yv$ have the same marginal and covariance but different joint distributions (as targeted).

Using the same construction procedure, other pairs of examples with the same marginals, same (non $\delta$-) autocovariance functions but different joint distributions could as easily be devised.
Mixture of Gaussians are used here by convenience, but mixtures of any other valid distributions could just as easily be reached.

\section{Conclusion and Perspectives}
Inspired from \emph{Statistical Physics} models, a general framework enabling us to define the joint distributions of a random vector $\Xv$ has been described.
It has then been specified to the definition of stationary time series, with control of their joint distributions and explicit derivation of numerous of their statistical properties.
A remapping onto a Hidden Markov Model enabled us to devise an efficient synthesis procedure, available upon request (in {\sc Matlab}). 
Constructive examples aiming at showing the potential of the tools were proposed. 
This general framework will be further explored by departing from the restrictive choices made here for the matrices $A$ and $\E$. 
This should notably enable us to define (and synthesize numerically) vectors of dependent variables with different marginal distributions and intricate dependencies. 
Notably, the intriguing and promising special cases of non-diagonalisable matrices $\E$ are expected to offer a much larger versatility in the form of dependencies that can actually be reached.
This is under current investigations. \\

J.-Y. Tourneret and N. Dobigeon are gratefully acknowledged for fruitful discussions.

\bibliographystyle{IEEEbib}
\vspace{-0.25cm}
\nopagebreak[4]
\bibliography{ProbaMat,ASEP}

\end{document}